# Analysis of Location Management Schemes for MANET using Synthetic Mobility Models

Mrs. H. A. Bhute
Pursuing M. E. Computer Engg.
(Computer Networks)
S.C.O.E., Vadgaon (Bk.)Pune-41

Prof. G.T. Chavan
Asst. Professor
Department of Comp. Engg
S.C.O.E., Vadgaon (Bk.) Pune-41

Prof. A.N. Bhute
Asst. Professor
Department of IT
S.C.O.E., Vadgaon (Bk.) Pune-41

**ABSTRACT**
In the performance evaluation of a protocol for an ad hoc network, the protocol should be tested under realistic conditions including, but not limited to, a sensible transmission range, limited buffer space for the storage of messages, representative data traffic models, and realistic movements of the mobile users (i.e., a mobility model). This paper is a survey of mobility models that are used in the simulations of ad hoc networks. We describe several mobility models that represent mobile nodes whose movements are independent of each other (i.e., entity mobility models) and several mobility models that represent mobile nodes whose movements are dependent on each other (i.e., group mobility models ).The goal of this paper is to simulate the movements of mobile nodes within a network and present a number of mobility models in order to demonstrate its effect on Location management scheme for mobile ad hoc network or personal communication services networks. Specifically, to illustrate how the performance results of an ad hoc network protocol drastically change as a result of changing the mobility model simulated.
Location management is a fundamental problem in personal communication services network. The current standard of location management is HLR/VLR scheme. It has been observed that the performance of any location management scheme depends on space requirements, bandwidth requirements and time requirements. To avoid certain drawbacks in HLR/VLR scheme, many approaches including hierarchical approaches have been suggested. Working set idea is chosen to analyze its performance for location management in PCS networks. Due to inadequacy of standard network simulators to provide the output in the format desired, a new location management simulator can be built. Two variants of working set idea viz. Working set scheme for HLR/VLR approach and working set scheme for hierarchical approach can be used and then compare the performance of HLR/VLR scheme and working set scheme using the results obtained by the simulator with respect to already available mobile activity traces. Working set scheme can also be analyzed for hierarchical networks. It is indicated that the results of working set idea reduces call setup time significantly at the expense of minor increase in database and link capacities in the personal communication services network from the literature survey.
*Keyword: Location Management, Mobility Models, Synthetic Mobility Models, Working set.*

## I. INTRODUCTION

Location management [4] refers to accessing and maintaining user information for call routing purposes. Important per user information, such as current location, authentication information and billing information are stored in user profiles. From an operational perspective, location management relies on two functions, profile lookups and profile updates. A profile lookup occurs in any call between users

- To access the callers profile for authentication, and
- To access the callee's profile for location information and connection status.

A profile update occurs
- To signal user equipment activation or deactivation
- To signal user call connection, or
- To register user movement.

Whenever a call needs to be delivered to the mobile, the network uses the last known location of the mobile terminal to search for the mobile in the vicinity of that area. This may involve paging for the mobile terminal in certain neighborhood of the last known location of the mobile terminal.

### A. INTRODUCTION TO THE PROBLEM OF LOCATION MANAGEMENT

Issues involved in any location management strategy are:

- Should the location information be centralized or distributed?
- If it should be distributed, where and how should this information be placed?
- How should an update or query operation be directed in a distributed system?
- What costs are involved in terms of resources and how do we optimize?
- How should the fault tolerance be achieved and what should be level of fault tolerance?
- How to minimize lookup and search cost for optimal search?
- Based on which parameters analysis is to be done?
- Is the approach suggested scalable?
- What should be the underlying topology for effective implementation of suggested approach?
- Does the scheme achieve fault tolerance?

The performance of any location management strategy is measured in the following terms.

**Space requirements**: Space requirements are the measure of how much space is required to store the location information of mobile host. Location registers (LRs) in the PCS network are responsible to store location information of mobile host. Update of location information causes data to be written at location register. Deregistration information causes location information to be removed from database. Both update and deregistration operations and lookup operations require access to the database. If frequency of these operations is not supported by the storage device for the particular scheme, then the scheme is not suitable for that network.

**Network bandwidth requirements** Network bandwidth requirements is the measure of how much network bandwidth is used to send data across the network. For the problem of location management, this data is registration, deregistration and lookup messages sent over the network. Each message has some bandwidth associated with it which depends upon the implementation of protocols related to sending these messages. If the bandwidth utilization by these messages is not supported by the network lines, then the scheme is not suitable for that network.

**Time requirements:** Time complexity is the measure of the time taken to get the location information of mobile host. Location information of mobile host is to be obtained within certain time bounds. This time taken is dependent on the cost of the links of the network and number of hops taken by the update, deregistration and lookup messages sent over the network. If the time taken to get the location of mobile host is exceeding some limit imposed by the users, then the particular scheme is not suitable for that network.

One strategy used in conventional systems to balance the cost of update and search is the use of registration area (RA) approach to location tracking. The geographical area is divided into several registration areas, where each registration area consists of several cells. The system tracks a mobile terminal registration area instead of its cell. Whenever a mobile terminal crosses from one registration area to another it informs its new location to the system. To setup a call to a mobile terminal, the system pages all the cells in the registration area to find the current cell of the mobile terminal. A database called location register (LR) is associated with each registration area to keep information about the mobiles currently registered in that registration area. [15].

### B. LOCATION MANAGEMENT APPROACHES

There are various approaches suggested to solve the problem of location management in mobile networks which can be broadly classified on the basis of storing user profile as

**Centralized approach:** Here user information is kept on only one node in the mobile network. For example, existing location management standards, IS41 and GSM are centralized approaches to location management [2]. Profile lookup and update operations are simple in this case but due to severe problems like congestion, central point failure etc. other approaches have been suggested.

**Distributed approach:** Here user information is distributed among many nodes in the network. Profile lookup and update operations are somewhat complex in this case but many schemes have come forward to solve this problem. For example, location management scheme proposed in [6] uses read and write set of location registers for lookup and replication of user information at more that one database.

**Hierarchical approach:** Hierarchical approach is a specialized version of distributed approach. Hierarchical approaches have been suggested to overcome the drawbacks of centralized approach. In this approach, location registers in the PCS network are arranged in hierarchical fashion. The topology resembles to a tree with a root level LR and its ascendant LRs while each leaf level LR performs location management operation for one zone of PCS network. Various approaches have been suggested regarding how to replicate user profile in the hierarchy. For example, location management scheme described in[2] replicates user information in all LRs in the hierarchy. One key problem in hierarchical approach is what should be the topology of hierarchy. The problem of optimal placement of location management directories in the hierarchy is solved using dynamic programming algorithm.

### C. MODELING LOCATION MANAGEMENT IN PCS NETWORKS

Previous studies have shown that for projected number of PCS users, existing location management standards, IS-41 and GSM will incur a large increase in database loads over the current levels. Actual performance of the suggested location management techniques depends strongly upon user behavior. As a result, realistic user behavior models are critical aspects in performance evaluation. Modeling location management addresses issues such as what should be topology model, call model and movement model. Components of modeling location management are described below.

**Basic Topology model:** The basic topology model is composed of the following objects:

- **User** represents a human user. A user's object contains information describing the user's current geographical location, the user's home location and the database(s) currently containing a copy of the user profile.

- **Site** representation geographical area. All site objects together define the physical geography for user movements. A site usually corresponds to the area covered by one profile database.

- **Database** represents any form of user database. A database is often associated with a site. Each database object maintains access statistics relating to number of read and writes, database messages sent, and total cost of sending all database messages (e.g. in hop counts).

- **Link** represents a direct communication link between two databases. It has a link cost describing the cost of sending a message through it. It maintains traffic statistics in terms of number of messages.

- **Geographical topology** is defined by a movement connectivity matrix which specifies for each site, its neighbors and the probabilities of users crossing into each of them. Network topology for communications between databases is specified though the links connecting them.

**Call model:** The call models describe how often individual users place calls to other people and characterizes the behavior of each call and how the callee is generated for each call.

**Movement model:** Movement model characterizes user movements within the geography defined by the basic topology model.

Section II describes the mobility models used for MANET. Section III describes the features of location management schemes. Section IV describes the working set scheme for Location Management.

## II. MOBILITY MODELS

Ad-hoc networks are a consequence of the ceaseless research efforts in mobile and wireless networks. They are a new paradigm of wireless communications for mobile hosts that are resource-constrained with only limited energy, computing power and memory. Each Mobile Ad hoc Network is a collection of wireless mobile hosts forming a temporary network without the aid of any established infrastructure or centralized administration, where each node acts both as a router and as a host. They are applied most commonly in situations such as military and emergency operations, target tracking, law enforcement, and rescue missions during disaster, etc. The definition of realistic mobility models is one of the most critical and, at the same time, difficult aspects of the simulations of networks designed for real mobile ad hoc environments. The reason for this is that most scenarios for which ad hoc networks are used have features such as dynamicity and extreme uncertainties (e.g. disaster). Thus use of real life measurements is currently almost impossible and most certainly expensive. Hence the commonly used alternative is to simulate the movement patterns. Most of the simulations use Random Waypoint model and its variants as they are designed to emulate movement of mobile nodes in a simplified fashion.

In order to evaluate the performance of a new protocol for a mobile ad hoc network, it is imperative to use a mobility model that accurately represents the mobile nodes (MNs) that will eventually utilize the given protocol. Only in this type of scenario it is possible to determine whether or not the proposed protocol will be useful when implemented. Currently there are two types of mobility models used in the simulation of networks: Traces and Synthetic models. Traces are those mobility patterns that are observed in real life systems. Traces provide accurate information, especially when they involve a large number of participants and an appropriately long observation period. However, new network environments e.g. MANETs are not easily modeled if traces have not yet been created. In this type of situation it is necessary to use synthetic models [1].

Synthetic models attempt to realistically represent the behaviors of MNs without the use of traces. This seminar presents several synthetic mobility models that have been proposed for the performance evaluation of ad hoc network protocols. A mobility model should attempt to mimic the movements of real MNs changes in speed and direction must occur and they must occur in reasonable time slots. For example, one would not want MNs to travel in straight lines at constant speeds throughout the course of the entire simulation because real MNs would not travel in such a restricted manner.

Seven different synthetic entity mobility models for ad hoc networks are discussed:
1. Random Walk Mobility Model: A simple mobility model based on random directions and speeds.
2. Random Waypoint Mobility Model: A model that includes pause times between changes in destination and speed.
3. Random Direction Mobility Model: A model that forces MNs to travel to the edge of the simulation area before changing direction and speed.
4. A Boundless Simulation Area Mobility Model: A model that converts a 2D rectangular simulation area into a torus-shaped simulation area.
5. Gauss-Markov Mobility Model: A model that uses one tuning parameter to vary the degree of randomness in the mobility pattern.
6. A Probabilistic Version of the Random Walk Mobility Model: A model that utilizes a set of probabilities to determine the next position of an MN.
7. City Section Mobility Model: A simulation area that represents streets within a city.

There are other synthetic entity mobility models available for the performance evaluation of a protocol in a cellular network or personal communication system (PCS). Although some of these mobility models could be adapted to an ad hoc network, this paper focuses on those models that have been proposed for (or used in) the performance evaluation of an ad hoc network.

Five group mobility models that allow researchers to simulate situations are presented; the MNs' decisions on movement depend upon the other MNs in the group.

- Exponential Correlated Random Mobility Model: A group mobility model that uses a motion function to create movements.
- Column Mobility Model: A group mobility model where the set of MNs form a line and are uniformly moving forward in a particular direction.
- Nomadic Community Mobility Model: A group mobility model where a set of MNs move together from one location to another.
- Pursue Mobility Model: A group mobility model where a set of MNs follow a given target.
- Reference Point Group Mobility Model: A group mobility model where group movements are based upon the path traveled by a logical center.

In all five group mobility models [1], random motion of each individual MN within a given group occurs. It is illustrated that a mobility model has a large effect on the performance evaluation of a working scheme for location management in mobile ad hoc network. In other words, it shows how the performance results of an ad hoc network protocol significantly change when the mobility model in the simulation is changed. The results presented prove the importance of choosing an appropriate mobility model (or models) for a given performance evaluation. The survey of number of synthetic mobility models used in ad hoc network simulations is done. The details of the models provide a good resource to researchers when they are deciding upon a mobility model to use in their performance evaluations.

## III. FEATURES OF LOCATION MANAGEMENT SCHEMES

Various schemes have been described to overcome drawbacks of existing standard for location management in PCS networks. The schemes discussed in survey differ in many aspects such as their nature, analysis, system models, information lookup and update policies etc. The features of these schemes can be summarized on the basis of following points:

**Scalability**

Scalability refers to the issue that if the proposed scheme is suitable for small areas as well as large areas spanning over the world.

It can be seen clearly that centralized schemes are not scalable. HLR storing information of MH can be distributed over a large geographical distance, thus increasing call setup time. Also, HLR is also subject to central point of failure and thus can become bottleneck in the system. Analyses of some schemes have been done for grid based N x N topology. But actual topology of mobile network is not grid based. Instead it is the function of geographic pattern of that region. Assumption that the topology is grid based simplifies analysis theoretically, but the results obtained can not be applied for day-to-day network topology. Also, grid based schemes are scalable in a limited sense, as the grid size can not be scaled both length and breadth wise for very large geographical area which can not be fitted into square region. Hierarchical schemes are scalable in the view that by increasing or decreasing the level of hierarchy which in turn depends upon the number of zones (location areas) in the network can be scaled for appropriate area.

**Fault Tolerance**

Fault tolerance of the system can be achieved by adding extra resources to it. Centralized schemes can not tolerate failure of HLR. If HLR fails, then MHs registered with it can not be serviced or contacted till the HLR recovers. Distributed schemes avoid this drawback by replicating location information across the network. But still the MH can not be contacted if the LR serving it is subject to failure. MH can be next contacted when it moves to another region serviced by another LR. When the failed LR recovers, then it need to recollect information about MHs it is serving. Communication links are also subject to failure. If the link connecting two regions fails, then MHs in those regions can no longer be able to communicate to each other until the link recovers. Many schemes described above assume stable hardware. In case of some fault, they will not be able to work properly.

**Resource Requirements**

Resources being scarce, performance of location management scheme depends on the resource consumption. Schemes such as [8] assume no restriction on bandwidth. So assumption in [8] is not appropriate if the scheme is to be applied for day to day networks. In the case of hierarchical networks, as the number of levels in the hierarchy of location registers increases, the load in the upper level of location registers (location directories) also increases. As the size of hierarchy grows, the time required to obtain location of mobile host also increases. So, limit on the number of levels of hierarchy is required. Better utilization of bandwidth required to eliminate unnecessary updates and deregistration messages.

## IV. IMPLEMENTATION OF WORKING SET SCHEME FOR LOCATION MANAGEMENT

Concept of working set [9] is based on the characteristics of MH that the set of sources that a given MH communicates most frequently with is very small. This set is also relatively stationary. Experiment was conducted on users to keep track of emails they receive. Results of experiment on two users are shown in Figure 2.4. From the statistics, it is clear that users have received most of the mails from few sources. This set of sources is termed as working set for a MH. Location information of MH is replicated at the site where the element of working set has registered is done if replicating information of MH incurs less overhead than searching for the MH.

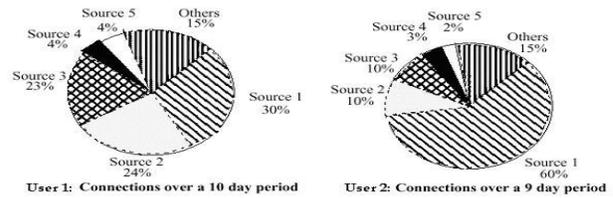

Figure 1. Statistics of emails received by two persons from various

Here, the working set for MH is calculated by MH. Once working set is determined, minimum spanning tree is calculated to disseminate location information to appropriate sites. It is based on the characteristics of mobile host that the set of sources that a given mobile host communicates most frequently with is very small. This set also called as working set for a mobile host. Location information of mobile host is replicated at the site where the element of working set has registered if replicating information of mobile host incurs less overhead than searching for mobile host. In the proposed scheme, the working set for mobile host is calculated by mobile host. Once working set is determined, location information of mobile host is passed to the elements in the working set. The variant of existent idea of working set to make it suitable for location management in PCS networks. The only difference between the original strategy and variant of strategy used by us is that in the modified strategy, the elements of the working set are zones instead of mobile hosts in[9]. The zones are static and information can be easily replicated in location registers in the respective zone. The other entities involved are accordingly mapped to corresponding entities in PCS network.

**The Adaptive Location Management Scheme**

The issues facing the adaptive scheme in [9] are, how to determine the current working set of the mobile host and how to determine which sources in the working set must be updated as the mobile host moves. The adaptive scheme uses an online algorithm that resolves both these issues together. For each source that communicates with mobile host, the adaptive scheme evaluates whether overall cost could be reduced by allowing the source to do a search, each time it sets up a connection to mobile host, or by updating the source each time the mobile host moves. This evaluation is done by computing the following quantities for each source s that actively communicates with the mobile host:

- $f_s$ : The estimated frequency with which connection set-up requests are received from the source s.

- $\delta_s$ : The additional cost incurred by s, in setting up a connection request to the mobile host. This cost is defined as
cost(s, HA) + cost(HA, MH) – cost (s, MH)………….. I
where HA = home agent in mobile IP protocol.
In short $\delta_s$ is the additional overhead incurred when the source does not know the current location of mobile host.

- $f_{update}$ : The estimated frequency with which mobile host changes its location and

- $U_s$: The cost incurred in both, announcing the location of mobile host to the source s and also invalidating the location information at source. Invalidating location information means that the location information at the source is made obsolete and all future connection requests from the source directed to the mobile host have to be routed through HA.

Having determined these parameters, the adaptive scheme [3] evaluates in an online manner whether the following inequality holds for the sources:

$$f_s * \delta_s > f_{update} * U_s \qquad \ldots\ldots\ldots\ldots..II$$

The left and right hand sides of the inequality denote routing cost and the update cost components, respectively. The above parameters are updated and the inequality is evaluated at the mobile host each time the source sets up a connection or when the mobile host moves. More details of this scheme can be found in [9].

**Variant Of Adaptive Scheme For HLR/VLR Approach**

The idea behind transferring adaptive scheme for HLR/VLR approach for PCS networks is that mobile users in PCS networks tend to be highly mobile. So, if the location information of mobile host is replicated to the location where mobile host in its working set is roaming and before next call, that mobile host moves out of its current zone then one won't get the benefit of replicating location information. So, instead of adding individual mobile users to working set, zones are added to working set which are stationary and if the location information of mobile host is to be replicated at that zone or not is decided based on modifications in parameters described as:

- **$f_s$**: The estimated frequency with which connection set-up requests are received from the mobile users in source s where s is a zone in network.

- **$\delta_s$**: The additional cost incurred by s, in setting up a connection request to the mobile host. This cost is defined as cost(s, HLR) + cost(HLR, CalleeZone) – cost (s, calleeZone)  …………….III
  where CalleeZone = zone where the callee mobile host is located and cost(s, HLR) + cost(HLR, CalleeZone) represent cost incurred in setting up connection request in HLR/VLR scheme. In short $\delta_s$ is the additional overhead incurred when location register in the zone from where call is done does not have the current location information of callee.

- **$f_{update}$**: The estimated frequency with which mobile host changes its location and

- **$U_s$:** The cost incurred in both, announcing the location of mobile host to the source s and also invalidating the location information at source. Invalidating means that the information at the source is made obsolete and all future connection requests from the source directed to the mobile host have to be routed through HLR.

The inequality equation remains same as equation II. The above parameters are updated and the inequality is evaluated at the mobile host each time the source sets up a connection for that source or when the mobile host moves for all possible sources. Here, HLR is the home location register of callee. This variant of adaptive scheme for HLR/VLR approach will be referred to as working set scheme for HLR/VLR approach in the subsequent chapters.

**Structure Of Hierarchy H Used In Experimental Analysis Of Working Set Scheme**

Conceptual diagram showing arrangement of location registers is shown in Figure 2

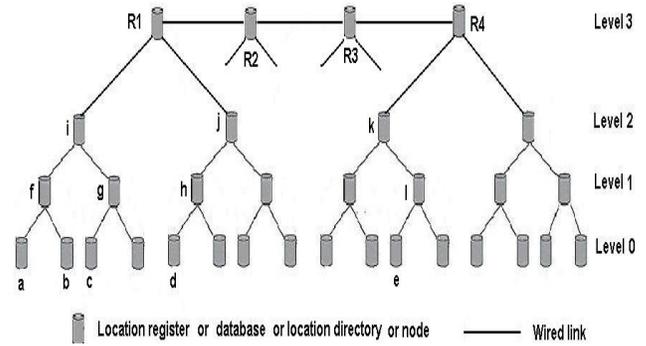

Figure 2 .Conceptual diagram showing hierarchical arrangement of databases

In H, it is assumed that location register and database performing location information related operation also called as location directory is a single entity. Thus there is database in each zone of the network which is shown at level 0 in the hierarchy shown in Figure 2. It is also assumed that for each location register (node of the tree) at upper level of the hierarchy tree, there are at least two children of that node. There can be one or more than one nodes at the root level. In the Figure 2, sub trees of root level directories R2 and R3 have not shown. All root level databases are connected by a single wired link as in bus topology. It's also assumed that all wired links in the network are bidirectional. User movements take place in zones of the network each of which has a location directory as shown in leaf level of hierarchy. Databases in upper level of hierarchy have a pointer to its child if the user is in the leaf level zone of that database. From the hierarchy structure, it is clear that there is a single and unique path from each node in tree to each other node in the tree.

**Location Update and Deregistration Operations In Hierarchy**

Consider movement of a user U from zone a to zone b shown in Figure 2. In this case, least common ancestor (LCA) of a and b which is f is updated with location information. Database in zone b is updated with information that user U is in zone b. this information is sent to database f which sends deregistration message to a informing that user U is not in zone a. Database in zone a deletes entry about location of user U. If a user U moves from zone a to zone c then databases g and i are sent update messages. After receiving update message, g sends deregistration message to databases f and a. If a user U moves from zone a to zone d then databases h, g and R1 are sent update messages. After receiving update message, R1 sends deregistration message to databases i, f and a. If a user U moves from zone a to zone e then databases l, k and R4 are sent update messages. R4 sends update messages to all root level directories that user U is in the sub tree of root level node R4. After receiving update message, R1 sends deregistration message to databases i, f and a. The update and deregistration messages are not multicast messages. The update message passes in a hop by hop fashion till the LCA of the zones in which there is a movement. Deregistration message is passed downwards the network till it reaches the old location of user U. Thus, for movement between zone a and zone d, update message sent by d follows path d – h – j – R1 which is terminated at R1 which is LCA of a and d. R1 then sends deregistration message to a which follows path R1 – i – f – a. For movement between zone a and zone e, update message sent by e follows path e – l – k – R4 which is terminated at R4. There is no LCA of a and e. So, R4 sends update message to all root nodes which follows path R4 – R3 – R2 – R1. R1 then sends deregistration message to which follows path R1 – i – f – a. If there is movement between zones a and some zone which is leaf of sub tree of R3, then R3 sends update message in both directions. Thus, in right direction, it will follow path R3 – R4 and in left direction it follows path R3 – R2 – R1. Thus, in this hierarchy, each node in root level contains information about the location of each user in the network and point to the root level node in the hierarchy whose sub tree is serving the corresponding user. This scheme, though increases number of update messages in the network, reduces number of lookups in the network. Lookup operation in the hierarchy is described in following section.

**Lookup Operations In Hierarchy**

If the caller is in same zone as the callee (to which call is made) call is directly routed to the callee. In this case, lookup takes place at database serving that zone. If the callee is not located in the zone of caller, then if there is a least common ancestor of caller zone and callee zone, then query propagates from caller zone to LCA and from LCA to callee zone. If the callee is not located in the zone of caller, then if there is no least common ancestor of caller zone and callee zone, then query propagates from caller zone to its root database which knows the root database in whose sub tree, callee is located. So, lookup message is sent to that root database which sends lookup messages to the callee zone. For example, if caller is in zone a and callee is in zone e, then database at a faces lookup. Since callee is not located in zone a, lookup message follows path to root database of a which is R1 as a –f – i – R1. Since R1 has information that callee is located in sub tree of root database R4, further lookup message follows path R4 - k – l – e. Since callee is in zone of database e, lookup ends here.

**Variant Of Adaptive Scheme For Hierarchical Approach**

In the case of hierarchy structure, there is no concept of HLR and VLR. If a user U is in zone a of hierarchy tree shown in Figure 2, then only parent of database at zone a that is database f contains a pointer to the exact location of user U. All other databases containing location information of user U, point to some other database. Also, there is a unique path from each node in the tree to each other node in the tree. Each lookup message between two zones at the leaf level in the hierarchy follows a unique path. So, the set of nodes to be considered for working set scheme differs for hierarchical scheme than that for HLR/VLR scheme. In the case of hierarchical approach, consider all the other nodes in the hierarchy except the leaf level node say i in which user U is located and the parent of node i as both contain the information that user U is served by the leaf level node i. If the location information of mobile user U is to be replicated at the zone considered for working set or not is decided based on modifications in parameters as:

- $f_s$ : The estimated frequency with which lookup operations are done for obtaining location information of mobile user U at source s where s is a zone in network which is considered for working set of user U.

- $\delta_s$ : The additional cost incurred by s, in obtaining exact location of mobile host. This cost is defined as

cost(s, CalleeZone) – 2 ……………...….IV

where CalleeZone = zone where the callee mobile host is located and cost(s, CalleeZone) represent number of lookup messages required to find exact location of mobile host. The second entity is always two as if exact location information is replicated at source s, and then it will cost only two lookup messages. One lookup at source s and another lookup at CalleeZone. In short $\delta_s$ is the additional overhead incurred in terms of number of lookups when location register in the zone where lookup is done does not have the current location information of callee.

- $f_{update}$ : The estimated frequency with which mobile host changes its location and

- $U_s$ : The cost incurred in both, announcing the location of mobile host to the source s and also invalidating the location information at source. This cost is measured in terms of number of hops. Invalidating means that the information at the source is made obsolete and all future lookups take place.

The inequality equation remains same as equation II. The above parameters are updated and the inequality is evaluated at the mobile host each time for the sources which face lookup in the hierarchical scheme when mobile host receives a call or when the mobile host moves for all possible sources. In this case, call is said to be served locally only if the zone of caller contains exact location information of callee. This variant of adaptive scheme for HLR/VLR approach will be referred to as working set scheme for HLR/VLR approach in the subsequent chapters. Thus, variants of working set approach in[9] to be suitable and applicable to existing HLR/VLR scheme and hierarchical scheme.

**V. DISCUSSION AND CONCLUSION**

After theoretically analyzing the working set scheme for synthetic mobility models, it can be seen that Working Set for HLR/VLR outperforms HLR/VLR scheme in terms of the overall database, message and hops requirements as CMR increases. It is expected that Irrespective of the value of CMR, the working set scheme shows significant improvement in the lookup cost and local to overall lookup ratio. Experimental performance evaluation of working set scheme using optimal database hierarchies and traces modeling real life call and movement pattern can be done.